# CONSERVATIVE STATISTICAL POST-ELECTION AUDITS

By Philip B. Stark

*University of California, Berkeley*

There are many sources of error in counting votes: the apparent winner might not be the rightful winner. Hand tallies of the votes in a random sample of precincts can be used to test the hypothesis that a full manual recount would find a different outcome. This paper develops a conservative sequential test based on the vote-counting errors found in a hand tally of a simple or stratified random sample of precincts. The procedure includes a natural escalation: If the hypothesis that the apparent outcome is incorrect is not rejected at stage $s$, more precincts are audited. Eventually, either the hypothesis is rejected—and the apparent outcome is confirmed—or all precincts have been audited and the true outcome is known. The test uses a priori bounds on the overstatement of the margin that could result from error in each precinct. Such bounds can be derived from the reported counts in each precinct and upper bounds on the number of votes cast in each precinct. The test allows errors in different precincts to be treated differently to reflect voting technology or precinct sizes. It is not optimal, but it is conservative: the chance of erroneously confirming the outcome of a contest if a full manual recount would show a different outcome is no larger than the nominal significance level. The approach also gives a conservative $P$-value for the hypothesis that a full manual recount would find a different outcome, given the errors found in a fixed size sample. This is illustrated with two contests from November, 2006: the U.S. Senate race in Minnesota and a school board race for the Sausalito Marin City School District in California, a small contest in which voters could vote for up to three candidates.

**1. Introduction.** Votes can be miscounted because of human error (by voters or election workers), hardware or software "bugs" or deliberate fraud. Post-election audits—manual tallies of votes in individual precincts—are intended to detect miscount, especially miscount large enough to alter the outcome of the election.[1] To the best of my knowledge, eighteen states require or

---

[1]Post-election audits can also reveal process problems, programming errors, equipment malfunctions and other issues that should be addressed even if they do not change the

allow post-election audits [National Association of Secretaries of State (2007) and Verified Voting Foundation (2007)]. California is one. Since 1965, California Elections Code has required a hand count of the ballots in a random sample of 1% of the precincts in each county, plus one precinct for each contest not represented in the 1% sample.[2] A post-election audit of 1% of precincts is a reasonable check for gross error and malfunction. However, to provide high confidence[3] that a full manual recount would confirm the apparent outcome requires auditing a number of precincts that depends on the number of precincts in the contest, the number of ballots cast in each precinct, the apparent margin of victory and the discrepancies the audit finds. No flat percentage, short of 100%, gives high confidence in all circumstances.[4]

In August 2007, California Secretary of State Debra Bowen de-certified and conditionally re-certified electronic voting machines in California. One condition of re-certification is that elections be audited using a sample size that depends on "the apparent margin of victory, the number of precincts, the number of ballots cast in each precinct, and a desired confidence level that the winner of the election has been called correctly."[5] The method

---

outcome. And audits deter fraud. See Norden et al. (2007) and Jefferson et al. (2007). For more on election monitoring, see Bjornlund (2004). An alternative approach to detecting error and deterring fraud is the "quick count," which monitors the counting process at a random set of polling stations or precincts. See Estok, Nevitte and Cowan (2002). An advantage of quick counts is that they can monitor the process, not just the outcome. A disadvantage is that poll workers and potential fraudsters can know which precincts or polling places are being monitored before the counts are official. The U.S. Government Accountability Office has published many reports on the accuracy and reliability of voting systems and election outcomes [e.g., Elections: Federal efforts to improve security and reliability of electronic voting systems are under way, but key activities need to be completed (2005), Elections: The nation's evolving election system as reflected in the November 2004 general election (2006) and Hite (2007)].

[2]See, for example, California Elections Code §15360.

[3]The meaning of "confidence" in the election audit community differs from its meaning in statistics. The "confidence" that the apparent outcome is correct is 100% minus the $P$-value of the hypothesis that the apparent outcome differs from the outcome a full manual recount would find.

[4]Some audit laws, such as California's 1% law, use the same precinct sampling fraction for every contest in an election. The amount of error required to make the apparent outcome of a contest wrong depends on the margin in the contest. The probability distribution of the miscount an audit uncovers in a contest depends on how the sample is drawn and the sample size, and also on the number of precincts in the contest and the number of ballots and miscounted ballots in each contest in each precinct. And the amount of error required to produce to make one of the losing candidates appear to be the winner depends on the margin in the contest. Thus, the decision of whether to confirm an election outcome depends on variables that are specific to a single contest. The method developed here addresses one contest at a time.

[5]See www.sos.ca.gov/elections/elections_vsr.htm.

presented here solves that problem. I am not aware of any other method that does. New Jersey recently passed a bill that requires post-election audits of randomly selected precincts, "to ensure with at least 99% statistical power that for each federal, gubernatorial or other Statewide election held in the State, a 100% manual recount of the voter-verifiable paper records would not alter the electoral outcome reported by the audit. For each election held for State office, other than Governor and Lieutenant Governor, and for county and municipal elections held in 100 or more election districts (the procedure will) ensure with at least 90% statistical power that a 100% manual recount of the voter-verifiable paper records would not alter the electoral outcome reported by the audit."[6] Again, the method presented here is the only one I am aware of that meets this requirement.

The U.S. House of Representatives is considering a bill, H.R. 811, The Voter Confidence and Increased Accessibility Act of 2007 (Holt),[7] which requires post-election audits of federal elections. The sampling percentage depends on the apparent margin of victory. Because the sampling percentage does not take into account precinct sizes, the number of precincts in a contest or the errors uncovered during the audit, it does not guarantee any particular level of confidence that the apparent outcome agrees with the outcome a full manual recount would find.

The Massachusetts legislature is also considering a bill that would require post-election audits of 5% of precincts, H671. The bill demands a complete recount if the discrepancy between the manual count and the reported vote exceeds certain thresholds. Like H.R. 811, H671 requires sampling a percentage of precincts that does not depend on the number of precincts in the contest, so it does not guarantee any particular level of confidence that the correct candidate was named the winner—unless a full recount is triggered.

Minnesota has an audit law (SF 2743) that requires audits of elections for President, governor, U.S. Senator and U.S. Representative. The sample size in each county is related to the number of registered voters in the county, rather than the number of precincts in the county. The sampling percentage the law requires does not take into account the number of precincts in the contest or the margin, but it has provisions for increasing the sample size if discrepancies are found; large discrepancies can trigger a recount of a county or an entire congressional district. Like the bills mentioned above, the Minnesota audit law does not guarantee any particular level of confidence that the outcome of the election is correct. See also Section 5.2.

Previous papers on the statistics of post-election audits [e.g., Saltman (1975), McCarthy et al. (2008), Dopp and Stenger (2006) and Rivest (2006)] in essence have concentrated on the question, "if there is enough error overall to change the outcome of an election, how large a random sample of precincts must be drawn to have chance at least $1-\alpha$ of finding at least one error?"[8] If fewer precincts than that are audited, we will not have $1-\alpha$ confidence that the outcome of the election is correct, even if the audit finds no errors. That is because there are ways of distributing enough miscount to spoil the election that have chance greater than $\alpha$ of being missed entirely by the sample.

---

[6]www.njleg.state.nj.us/2006/Bills/AL07/349_.PDF.
[7]See holt.house.gov/HR_811.shtml.

If the sample is at least as large as these methods prescribe, and the manual tally finds no error, we are done: either the apparent winner is the true winner or an event with probability less than $\alpha$ occurred (or one of the assumptions of the method is wrong). But if the sample contains any miscount, however small, these approaches do not tell us how reliable the election outcome is, nor whether to confirm the outcome. The rules are incomplete.

Manual tallies routinely turn up small miscounts. What should we do then? Recount the entire contest by hand? Audit more precincts? If so, how many? What if the expanded audit finds more miscount? When do we stop? How do we decide whether the outcome is in doubt?

An audit procedure is incomplete unless it always either (i) confirms the outcome of the election or (ii) demands a full recount. And it should have an error rate that can be quantified in a reasonable way. For example, a procedure might come with a mathematical guarantee that if it confirms the outcome of the election, either the outcome is the same that a full manual recount would find, or an event with probability no greater than $\alpha$ occurred.[9]

Deciding whether to confirm the outcome of a contest can be viewed as testing a null hypothesis. The null hypothesis for election audits can be chosen in more than one way. For example, the null hypothesis could be "the outcome is right" or "the outcome is wrong." In the Neyman/Pearson paradigm, the chance of a type I error, the error of rejecting the null hypothesis when it is true, is controlled to be at most $\alpha$, the significance level.[10] The risk of incorrectly rejecting the null hypothesis when it is true is primary.

---

[8]The computations in those papers assume that the precincts to be hand-tallied are a random sample without replacement drawn from all the precincts in the contest. However, in California, the precincts for audit are not chosen that way. Rather, 1% of the precincts in each county are chosen at random (additional precincts are chosen, not necessarily at random, if contests are missed by the sample). This is a stratified random sample of precincts, not a simple random sample of precincts.

[9]See Section 6.5 for other possibilities.

[10]One can try to find the level-$\alpha$ test that maximizes the power, the chance of rejecting the null hypothesis when a particular alternative is true.

In election auditing, the primary risk is that of confirming an outcome that is wrong. Failing to confirm an outcome that is correct—on the basis of an initial audit sample—could lead to additional auditing, but that economic risk seems less serious than the risk of awarding the contest to the wrong candidate. We want the audit to provide strong evidence that the contest came out right, not just to fail to find evidence that the contest came out wrong. Hence, it makes sense to choose the null hypothesis to be that the outcome is wrong, and to devise a test that has probability at most $\alpha$ of incorrectly rejecting that hypothesis. If we reject the hypothesis that the outcome is wrong, we conclude that the apparent outcome is the outcome a full manual recount would find. If not, we count more votes. Eventually, either we confirm the outcome or we have recounted all the ballots by hand.

This paper constructs a conservative sequential test of the hypothesis that the apparent outcome is not the outcome a full manual recount would find. The test terminates either with the declaration that the apparent outcome is correct or with a full recount. The chance is at most $\alpha$ that the procedure declares that the outcome is correct if the outcome is not the outcome a full manual recount would find. The procedure also gives a $P$-value for the hypothesis that the outcome is incorrect: a number $P$ such that, given the errors observed in the sample, either a full manual recount would find the same outcome or an event that had probability no greater than $P$ occurred.

In the approach developed here, an audit can confirm the outcome of a contest, but only a full manual recount can invalidate the outcome. So, there is a positive probability that a declaration that the election outcome is correct is mistaken, but a declaration that the outcome is incorrect is as certain as a full manual recount can be. The approach automatically leads to a full recount if the outcome of the contest is not validated by a manual tally of some sufficiently large random sample of precincts.

There are many ad hoc choices in the method below, and the approach is not the most powerful [with a different method, it might be possible to get the same confidence by auditing fewer precincts. See, e.g., Stark (2008b).] The choices were made to simplify the exposition and implementation: methods need to be transparent to be adopted as part of the election process and to inspire public confidence. For example, an approach that required numerical optimization to maximize $P$-values for a likelihood ratio test statistic over sets of nuisance parameters might be more efficient, but because of its complexity would likely meet resistance from elections officials and voting rights groups. In contrast, the most esoteric calculation required for the method presented here is $\binom{N}{n}$. It could be implemented in a spreadsheet program, which is perhaps a good design criterion for software to be used by jurisdictional users at all levels of government.

The main point of this paper is not the method itself; rather, the method is an existence proof showing that it is possible to get conservative statistical measures of confidence in election outcomes from post-election audit results using simple computations.

TABLE 1
*Notation*

| | |
|---|---|
| $C$ | number of counties with at least one precinct in the contest. |
| $\mathcal{C}$ | the integers $\{1, \ldots, C\}$. |
| $N \equiv \sum_{c \in \mathcal{C}} N_c$ | number of precincts in the contest. |
| $\mathcal{N}$ | the integers $\{1, \ldots, N\}$. |
| $N_c$ | number of precincts in the contest in county $c$. |
| $\mathcal{J}_n^\star$ | a simple random sample of $n$ elements of $\mathcal{N}$. |
| $\mathcal{J}_n^\diamond$ | a random sample with replacement of $n$ elements of $\mathcal{N}$. |
| $b_p$ | reported voting opportunities in precinct $p$, $f$ times the number of ballots reported in precinct $p$, including undervoted and invalid ballots. |
| $B_c$ | reported voting opportunities in county $c$. |
| $B \equiv \sum_{p \in \mathcal{N}} b_p = \sum_{c \in \mathcal{C}} B_c$ | reported voting opportunities in the contest. |
| $K$ | number of candidates and pseudo-candidates in the contest, after pooling. See Section 3.1. |
| $\mathcal{K}$ | the integers $\{1, \ldots, K\}$. |
| $\mathcal{K}_w$ | the indices of the $f$ candidates who are apparent winners. |
| $\mathcal{K}_\ell$ | the indices of the $K - f$ candidates who are apparent losers. |
| $a_{kp}$ | actual vote for (pseudo-)candidate $k$ in precinct $p$. |
| $A_k \equiv \sum_{p \in \mathcal{N}} a_{kp}$ | actual total vote for (pseudo-)candidate $k$. |
| $r_p$ | upper bound on $\sum_{k \in \mathcal{K}} a_{kp}$, the actual total vote in precinct $p$. |
| $v_{kp}$ | reported vote for (pseudo-)candidate $k$ in precinct $p$. |
| $V_k \equiv \sum_{p \in \mathcal{N}} v_{kp}$ | total vote reported for (pseudo-)candidate $k$. |
| $M$ | overall apparent margin in votes: reported votes for the apparent winner(s) with fewest reported votes, minus reported votes for an the apparent loser(s) with the most reported votes: $M = \bigwedge_{k \in \mathcal{K}_w} V_k - \bigvee_{k \in \mathcal{K}_\ell} V_k$. |
| $e_p \equiv \sum_{k \in \mathcal{K}_w} (v_{1p} - a_{1p})_+$ $+ \sum_{k \in \mathcal{K}_\ell} (a_{kp} - v_{kp})_+$ | maximum by which error in precinct $p$ could increase $M$. |
| $u_p$ | a priori upper bound on $e_p$. See Section 3.2. |
| $E = \sum_{p \in \mathcal{N}} e_p$ | maximum by which error in all precincts could increase $M$. |
| $w_p(\cdot)$ | a monotonic weight function for error in precinct $p$. See Section 3.3. |
| $w_p^{-1}(\cdot)$ | the inverse of $w_p$: $w_p^{-1}(t) \equiv \sup_z\{z : w_p(z) \le t\}$. |

**2. Assumptions and notation.** Table 1 sets out the notation. All variables refer to a single contest of the form "vote for up to $f$ candidates." Each ballot has $f$ voting opportunities for the contest; there are $f$ apparent winners of the contest. A ballot with votes for more than $f$ candidates is *overvoted*.

Overvotes are invalid—they do not count as votes for any candidate.[11] A ballot with votes for fewer than $f$ candidates is *undervoted*. The number of undervotes on such a ballot is $f$ minus the number of votes.

The analysis uses the following assumptions:

1. All kinds of error are possible in the machine counts: there can be errors in the number of valid votes for each candidate, undervotes and invalid votes. Ballots can be overlooked entirely. Ballots that do not exist can be counted.
2. The truth is whatever the hand tally shows. (When the hand count does not match the machine count, the hand count is typically repeated until the counters are confident that the problem is with the machine count. Hand counts are subject to error, but they are the gold standard.)
3. Precincts are selected at random for post-election audit.

Among the apparent losers, any candidate with at least as many reported votes as the rest is an "apparent runner-up." The apparent margin $M$ is the difference between the number of votes reported for the apparent winner(s) with the fewest reported votes and the number of votes reported for an apparent runner-up. If more than $f$ candidates have at least as many reported votes as the apparent top $f$ candidates, $M = 0$: the contest is apparently tied for the last winning place.

More precisely, let $V_k$ be the total number of votes reported for candidate $k$, $k = 1, \ldots, K$. Let $(V_{(k)})_{k=1}^{K}$ be the votes $(V_k)_{k=1}^{K}$ in rank order, so that $V_{(1)} \geq V_{(2)} \geq \cdots \geq V_{(K)}$. Then $M = V_{(f)} - V_{(f+1)}$. If $\#\{k : V_k \geq V_{(f)}\} > f$, $M = 0$ and the contest is a tie.

As discussed in Section 3.1, some subsets of apparent losers (and undervotes and invalid ballots) can be pooled to form a smaller number of "pseudo-candidates." Pooling can reduce the sample size needed to confirm the election. After pooling, there remain $K$ candidates and pseudo-candidates, numbered 1 through $K$.

**3. Testing the election outcome.** The approach to testing whether the apparent election outcome is wrong is as follows:

1. Select a test statistic.[12]

---

[11] Some states have "voter intent" laws: the people conducting the hand tally try to determine what the voter intended, even if a machine could not. So, for example, a ballot that had a mark for George Washington and also had George Washington as a write-in candidate would be an overvote according to the machine, but a human might infer that the voter intended to vote for George Washington. This paper assumes that rules are in place for determining whether that is a valid vote.

[12] In principle, the choice could be optimized to maximize power against some alternatives. In practice, the method must be transparent, easy for the public to understand,

2. Select a sampling design and an increasing sequence of sample sizes $(n_s)$.[13] Select a corresponding sequence of significance levels $(\alpha_s)$ that give a level-$\alpha$ test overall.[14]
3. Set $s = 1$. Set the initial sample to be the empty set.
4. Augment the current sample by a random sample so that it contains $n_s$ precincts in all.
5. Tally the votes in the new precincts by hand.
6. Calculate the test statistic and the maximum $P$-value for the test statistic over all ways of allocating error among the precincts that would result in a different election outcome.
7. If the maximum $P$-value is less than $\alpha_s$, confirm the apparent outcome. Otherwise, increment $s$ and return to step 4, unless all $N$ precincts have now been hand tallied. If all precincts have been hand tallied, confirm the outcome the hand tally shows.

3.1. *Marginal notes.*

EXAMPLE 1. Consider a winner-take-all ($f = 1$) contest with $K = 2$ candidates. The reported vote for the apparent winner is $V_1 = 1{,}000$ votes, and the reported vote for the apparent loser is $V_2 = 500$ votes. The margin is $M = 1{,}000 - 500 = 500$ votes.

It is possible that both candidates actually had 750 votes and the apparent margin was produced by miscounting 250 ballots with votes for the apparent loser as votes for the apparent winner: if the apparent winner's vote total was high by 250 and the apparent loser's vote total was low by 250, that could have turned a tie into the apparent margin. Alternatively, if 500 apparent undervotes were miscounted as votes for the apparent winner, that could have turned a tie into the apparent margin. Or if 500 ballots with votes for the apparent loser had been overlooked on election day, that could have turned a tie into the apparent margin. Or if 500 ballots with votes for the apparent winner had been double-counted on election day, that could

---

easy for elections officials to implement, and easy to verify or replicate. Here I use the maximum of functions of the amount by which error in each precinct in the sample could have inflated the margin, after pooling subsets of losers as described in Section 3.1. This leads to simple probability calculations. See Section 3.3.

[13]We might increase the sample size by a fixed number of precincts at each stage, such as $\lceil 0.02N \rceil$. Or we might increment the sample by the smallest number of precincts such that, if the test statistic did not increase from its current value, we would confirm the outcome. The only requirement is that $n_{s+1} - n_s \geq 1$.

[14]For example, $\alpha_s \equiv \alpha/2^s$, $s = 1, \ldots$. Alternatively, if the sequence of sample sizes $(n_s)$ guarantees that by stage $S$ all $N$ precincts will be in the sample, we could take $\alpha_s = \alpha/S$. These choices just use Bonferroni's inequality; one could do better using methods from sequential analysis. See Section 6.4.

have turned a tie into the apparent margin. If 100 votes for the apparent loser had been miscounted as undervotes, 100 had been miscounted as votes for the apparent winner, 100 votes for the apparent loser had been overlooked entirely, and 100 nonexistent ballots had been counted as votes for the apparent winner, that could have turned a tie into the apparent margin. (Net, the reported vote totals would have been off by 200 for the apparent winner and 300 for the apparent loser, 500 votes in all.) But if the overcount for the apparent winner plus the undercount for the apparent loser is less than 500 votes in all, the apparent winner must be the true winner.

More generally, suppose there are $K$ candidates in all, undervotes, invalid ballots and overlooked ballots. (A negative number of ballots could be overlooked, corresponding to overcounting real ballots or counting nonexistent ballots.) An error that increases the count for any of the apparent winners by 1 vote increases the apparent margin by at most 1 vote. An error that decreases the count for any of the apparent losers by 1 vote increases the apparent margin by at most 1 vote. Conversely, errors that decrease the count for any apparent winner or that increase the count for any apparent loser might decrease the apparent margin, but cannot increase the apparent margin. Miscounting a vote for an apparent loser as a vote for an apparent winner could affect inflate the apparent margin by as much as 2 votes (or possibly 0 or 1). Miscounting an undervote as a vote for one of the apparent winners could increase the apparent margin by as much as 1 vote. Overlooking a valid vote for one of the losers could increase the apparent margin by as much as 1 vote. Errors in the number of undervotes or invalid ballots do not by themselves affect the margin.

In summary, the amount by which error could have artificially inflated the apparent margin is at most the total overcount for all the apparent winners, plus the total undercount for all the apparent losers.

Let $v_{kp}$ be the reported number of votes for candidate $k$ in precinct $p$, $V_k = \sum_{p \in \mathcal{N}} v_{kp}$ be the total number of reported votes for candidate $k$, $a_{kp}$ be the actual number of votes for candidate $k$ in precinct $p$, and $A_k = \sum_{p \in \mathcal{N}} a_{kp}$ be the actual total number of votes for candidate $k$. Let $\mathcal{K}_w$ denote the indices of the candidates who are apparent overall winners of the race (so $\#\mathcal{K}_w = f$) and let $\mathcal{K}_\ell$ denote the indices of the candidates who are apparent losers. For real $z$, define $z_+ \equiv z \vee 0$. The *potential margin overstatement in precinct $p$* is

$$(1) \qquad e_p \equiv \sum_{k \in \mathcal{K}_w} (v_{kp} - a_{kp})_+ + \sum_{k \in \mathcal{K}_\ell} (a_{kp} - v_{kp})_+.$$

The *total potential margin overstatement* is $E \equiv \sum_{p \in \mathcal{N}} e_p$. The *net potential margin overstatement in the election* is

$$(2) \qquad \mathcal{E} \equiv \sum_{k \in \mathcal{K}_w} (V_k - A_k)_+ + \sum_{k \in \mathcal{K}_\ell} (A_k - V_k)_+.$$

We know that

(3) $$M = \bigwedge_{k \in \mathcal{K}_w} V_k - \bigvee_{k \in \mathcal{K}_\ell} V_k.$$

Thus,

$$\bigwedge_{k \in \mathcal{K}_w} A_k - \bigvee_{k \in \mathcal{K}_\ell} A_k \geq \left( \bigwedge_{k \in \mathcal{K}_w} V_k - \bigvee_{k \in \mathcal{K}_w} (V_k - A_k)_+ \right)$$
$$- \left( \bigvee_{k \in \mathcal{K}_\ell} V_k + \bigvee_{k \in \mathcal{K}_\ell} (A_k - V_k)_+ \right)$$

(4) $$\geq \left( \bigwedge_{k \in \mathcal{K}_w} V_k - \sum_{k \in \mathcal{K}_w} (V_k - A_k)_+ \right)$$
$$- \left( \bigvee_{k \in \mathcal{K}_\ell} V_k + \sum_{k \in \mathcal{K}_\ell} (A_k - V_k)_+ \right)$$
$$= M - \sum_{k \in \mathcal{K}_w} (V_k - A_k)_+ - \sum_{k \in \mathcal{K}_\ell} (A_k - V_k)_+$$
$$= M - \mathcal{E}.$$

So, the apparent set of winners must be the true set of winners if

(5) $$\mathcal{E} < M.$$

By the triangle inequality,

$$\mathcal{E} \equiv \sum_{k \in \mathcal{K}_w} (V_k - A_k)_+ + \sum_{k \in \mathcal{K}_\ell} (A_k - V_k)_+$$
$$= \sum_{k \in \mathcal{K}_w} \left( \sum_{p \in \mathcal{N}} (v_{kp} - a_{kp}) \right)_+ + \sum_{k \in \mathcal{K}_\ell} \left( \sum_{p \in \mathcal{N}} (a_{kp} - v_{kp}) \right)_+$$

(6) $$\leq \sum_{k \in \mathcal{K}_w} \sum_{p \in \mathcal{N}} (v_{kp} - a_{kp})_+ + \sum_{k \in \mathcal{K}_\ell} \sum_{p \in \mathcal{N}} (a_{kp} - v_{kp})_+$$
$$= \sum_{p \in \mathcal{N}} \left( \sum_{k \in \mathcal{K}_w} (v_{kp} - a_{kp})_+ + \sum_{k \in \mathcal{K}_\ell} (a_{kp} - v_{kp})_+ \right)$$
$$= \sum_{p \in \mathcal{N}} e_p \equiv E.$$

Hence, the apparent outcome must be the same that a full manual recount would show if $E < M$. Our test is based on this condition. For a sharper sufficient condition, see Stark (2008b).

EXAMPLE 2. Consider a winner-take-all ($f = 1$) contest with $K = 4$ candidates. The reported vote totals are $V_1 = 800$ votes, $V_2 = 500$ votes, $V_3 = 150$ votes, and $V_4 = 50$ votes. The margin is $M = 800 - 500 = 300$ votes. The reported winner might not be the real winner if 150 votes for candidate 2 had been miscounted as votes for candidate 1, producing a net potential margin overstatement in the election of 300 votes; then candidates 1 and 2 might have been tied. Candidate 3 could not have been the winner unless the net potential margin overstatement in the election is more than 650 votes, and candidate 4 could not have been the winner unless the net potential margin overstatement in the election is more than 750 votes. The apparent winner must be the true winner if $E < M$.

EXAMPLE 3. What if, in Example 2, we pretend that candidates 3 and 4 are a single "pseudo-candidate" with $150 + 50 = 200$ reported votes? Then $K = 3$ (pseudo-)candidates, with $V_1 = 800$ votes, $V_2 = 500$ votes, and $V_3 = 200$ votes. Candidate 1 must be the true winner if the net potential margin overstatement in the election for candidate 1, candidate 2 and pseudo-candidate 3 is less than $M = 300$ votes. If pseudo-candidate 3 could not have been the winner, then neither the original candidate 3 nor the original candidate 4 could have been the winner, because the pseudo-candidate gets all the votes for both of them—at least as many votes as either gets separately. The apparent winner must be the true winner if $E < M$, with $E$ measured for the three pseudo-candidates who remain after pooling candidates 3 and 4.

Pooling candidates 3 and 4 into a single pseudo-candidate tends to result in a more powerful test, because $e_p$, the potential margin overstatement in precinct $p$, then ignores errors that do not change the number of votes for the pseudo-candidate, such as counting a vote for candidate 3 as a vote for candidate 4 or vice versa. Such errors cannot suffice to change the outcome of the election. For the outcome to be wrong, *in addition* to errors that redistribute votes among the candidates who are pooled together, it is necessary that $E \geq M$.

If we pool candidates 2 and 3 into a single pseudo-candidate with $500 + 150 = 650$ votes, the margin between the apparent winner and that pseudo-candidate is only 150 votes. Provided the total potential margin overstatement measured for candidate 1, the pseudo-candidate and candidate 4 is less than 150 votes, candidate 1 must be the real winner. The sufficient condition for the outcome to be right has changed: we need $E < 150 < M$. In effect, we need to test using a smaller margin, the margin between the winner and the pseudo-candidate—the runner-up after pooling. That could result in a less powerful test, so we will avoid it. We cannot pool candidates whose total vote is greater than or equal to the vote for any of the apparent winners, because then the outcome of the contest could be wrong even if $E = 0$.

EXAMPLE 4. Suppose that the contest allows votes for up to two out of three candidates, so $f = 2$ and $K = 3$. Suppose that a full hand recount would show that candidate 1 got 1,000 votes, candidate 2 got 500 votes, candidate 3 got 500 votes, there were 250 undervotes and there were 250 overvoted ballots. Let $M \leq 500$. Miscounting $M/2$ of the votes for candidate 3 as votes for candidate 2 would produce an apparent margin of $M$ between them, with net potential margin overstatement in the election of $M$. Miscounting $M$ of the overvoted ballots as one vote each for candidate 1 and candidate 2 would produce an apparent margin of $M$ votes between candidates 2 and 3, with net potential margin overstatement in the election of $2M$. Failing to count $M$ of the votes cast for candidate 3 would produce an apparent margin of $M$ votes between candidates 2 and 3, with a net potential margin overstatement in the election of $M$. The outcome must be correct if $E < M$.

EXAMPLE 5. Finally, consider an example with $f = 2$, undervotes and overvotes and pooling. There are four candidates on the ballot, plus two write-in candidates. The reported votes are as follows: 500 votes for the apparent overall winner; 400 votes for the apparent second-place winner; 300 votes for the apparent runner-up (the loser with the most votes); 100 votes for the apparent fourth-place candidate listed on the ballot; 5 votes for each of the two write-ins; 50 undervotes and 50 invalid ballots. The margin is $M = 400 - 300 = 100$ votes. If we pool the fourth-place candidate, the write-ins, the undervotes and $f$ times the overvotes into a single pseudo-candidate, that pseudo-candidate would have $100 + 5 + 5 + 50 + 2 \times 50 = 260$ votes, fewer than the runner-up. So, we can take $K = 4$ candidates, corresponding to the apparent overall winner, the apparent second winner, the runner-up and the pseudo-candidate. The outcome of the election cannot be wrong unless the net potential margin overstatement in the election, measured for those four (pseudo-)candidates, is at least $M = 100$ votes. The total potential margin overstatement $E$ would have to be greater than 140 votes for the pseudo-candidate to be one of the winners, and if the pseudo-candidate is not a winner, neither the apparent fourth-place candidate nor any of the write-ins could be winners. Hence, the outcome must be correct if $E < M$.

We shall adopt the following rule for pooling:

POOLING RULE. Pool the losers into groups so that no group has more votes than the runner-up, but the group with the fewest votes has as many votes as possible.

Other pooling rules make sense too, for example, "pool the losers into as few groups as possible such that no group has more votes than the runner-up." Any such pooling rule ignores many errors that by themselves cannot affect the outcome of the contest, but still the apparent winners must be the true winners if the total potential margin overstatement $E < M$. If a pseudo-candidate cannot be the winner, then neither can any of the real candidates who were pooled to form the pseudo-candidate. The value of $K$ is the number of candidates and pseudo-candidates that remain after pooling. It is not necessary to pool—the test developed below is conservative even without pooling—but pooling yields a more powerful test. However, compare with Stark (2008b).

3.2. *Bounding the potential margin overstatement in each precinct.* If the potential margin overstatement in individual precincts can be large compared to the margin, it will take a large sample to provide compelling evidence that $E < M$, because an outcome-changing error could hide in a small number of precincts. Miscount that could affect the outcome of an election is easier to detect if it must be spread over many precincts.

By how much can error in precinct $p$ inflate the apparent margin? We need an upper bound $u_p$ for the potential margin overstatement $e_p$ in precinct $p$. The smaller the values $u = (u_p)_{p=1}^N$ are, the larger the number of precincts that must be "tainted" to have $E > M$, and so the easier it is to detect an election-altering amount of error. If any number of ballots could be overlooked or overcounted on election day, there is no finite bound $u_p$ for $e_p$.

Some studies assume that if the discrepancy in any precinct exceeds, say, 40% of the votes reported in precinct $p$ [Saltman (1975) and McCarthy et al. (2008)] or 40% of the ballots reported in precinct $p$, including undervotes and invalid ballots [Dopp and Stenger (2006)], that would be detected even without an audit. (If votes on 20% of the ballots had been "flipped" to an apparent winner from an apparent loser, that would produce a potential margin overstatement $e_p$ of 40% of the ballots.) That is, the studies take $u_p = 0.4 b_p$. This could be reasonable in some circumstances, but it is hard to justify.

Suppose we know a number $r_p \geq 0$ so that the actual total vote satisfies $\sum_{k \in \mathcal{K}} a_{kp} \leq r_p$. For example, in precincts that use optical scan ballots, the total number of votes can be no larger than $f$ times the number of ballots delivered to the precinct, so that could serve as $r_p$. The number of votes cast in a precinct can be no larger than $f$ times the number of voters registered in a precinct, including same-day registrations (if the jurisdiction allows them), so that could serve as $r_p$. A count of signatures in a precinct pollbook, times $f$, might provide a value for $r_p$, although occasionally someone might vote without signing in. In some jurisdictions, elections officials check the number of voted, spoiled and unvoted ballots in every precinct against the number of ballots sent to and returned from the precinct. The number of voted ballots according to such an "accounting of ballots," times $f$, could serve as $r_p$.

If $\sum_{k \in \mathcal{K}} a_{kp} \leq r_p$, it is impossible for $e_p$ to exceed

$$
\begin{aligned}
(7) \quad e_p^+(r_p) &\equiv \max_{x \in \mathbb{R}^K \,:\, x \geq 0,\, \sum_{k \in \mathcal{K}} x_k \leq r_p} \left\{ \sum_{k \in \mathcal{K}_w} (v_{kp} - x_k)_+ + \sum_{k \in \mathcal{K}_\ell} (x_k - v_{kp})_+ \right\} \\
&= r_p + \sum_{k \in \mathcal{K}_w} v_{kp} - \bigwedge_{k \in \mathcal{K}_\ell} v_{kp}.
\end{aligned}
$$

These bounds suppose that every one of the $r_p$ possible valid votes in precinct $p$ might in fact have been a vote for the apparent loser $k \in \mathcal{K}_\ell$ with the fewest reported votes in precinct $p$. Let $e^+(r)$ denote the $N$-vector with components $(e^+(r))_p = e_p^+(r_p)$, $p \in \mathcal{N}$.

Note that if some apparent loser $k \in \mathcal{K}_\ell$ gets no votes in precinct $p$, $e_p^+(r_p)$ takes its maximum possible value, $r_p + \sum_{k \in \mathcal{K}_w} v_{kp}$; $e_p^+(r_p)$ gets smaller as the minimum number of votes any apparent loser gets in precinct $p$ gets bigger. The pooling rule in Section 3.1 tends to make $\bigwedge_{k \in \mathcal{K}_\ell} v_{kp}$ larger than it would be without pooling. This is another way pooling helps, especially in contests with write-in candidates, because often there are many precincts in which some write-in candidate receives no votes.

Henceforth, $u$ will be a vector of upper bounds for $e$. Whether $u$ is $e^+(r)$, $0.4b$ or some other bound does not matter for the rest of the mathematical development.

3.3. *The test statistic.* For any $x \in \mathbb{R}^N$ and $\mathcal{J} \subset \mathcal{N}$, define

$$
(8) \qquad \bigvee_{\mathcal{J}} x \equiv \bigvee_{p \in \mathcal{J}} x_p
$$

and

$$
(9) \qquad \sum_{\mathcal{J}} x \equiv \sum_{p \in \mathcal{J}} x_p.
$$

For $x, y \in \mathbb{R}^N$, define $x \wedge y$ to be the vector with components

$$
(10) \qquad (x \wedge y)_p = x_p \wedge y_p, \qquad p \in \mathcal{N},
$$

and $x \vee y$ to be the vector with components

$$
(11) \qquad (x \vee y)_p = x_p \vee y_p, \qquad p \in \mathcal{N}.
$$

Fix a set of monotonically increasing functions $w = (w_p(\cdot))_{p=1}^N$ and for $x \in \mathbb{R}^N$, define $w(x) \equiv (w_p(x_p))_{p=1}^N$. Let $\mathcal{J}_n^\star$ be a simple random sample of size $n$ from $\mathcal{N}$. The hypothesis test is based on the test statistic

$$
(12) \qquad \bigvee_{\mathcal{J}_n^\star} w(e).
$$

The functions $w = (w_p(\cdot))$ quantify our relative tolerance for errors in different precincts $p \in \mathcal{N}$. All choices of $w$ yield conservative tests, so $w$ can be chosen at will. For example, we might choose $w_p(z) = z$. Then every error that could increase the apparent margin gets the same weight. Or we might choose $w_p(z) = z/b_p$; then $\bigvee_{\mathcal{J}_n^\star} w(e)$ is the maximum potential margin overstatement relative to the reported number of voting opportunities in the precinct. We might choose $w_p(z) = z/u_p$; then $\bigvee_{\mathcal{J}_n^\star} w(e)$ is the maximum potential overstatement of the margin as a fraction of the bound on the margin overstatement in the precinct. Or we might pick $w_p(z)$ to reflect the accuracy of the voting technology. For example, we might be less tolerant of error in precincts with direct-recording electronic (DRE) machines than we are of error in precincts with optically scanned ballots.[15] Then we might pick $w_p(\cdot)$ to grow more rapidly for DRE precincts than for precincts that use optically scanned ballots. Because post-election audits often find a miscounted vote or two, even in precincts with very few votes, weight functions of the following form can be desirable:

$$w_p(z) = (z-m)_+/b_p, \tag{13}$$

with $m$ on the order of 2 or 3. This function ignores potential margin overstatements of up to $m$ votes per precinct, and penalizes larger potential margin overstatements in inverse proportion to the size of the precinct (here size is the reported number of voting opportunities). That prevents an error in scanning a single ballot in a small precinct from making the test statistic large, but takes into account the fact that we expect more discrepancies in larger precincts, all other things being equal.

3.4. *Tail probabilities for the sample maximum.* This section shows how to find $P$-values for the hypothesis $E \geq M$ using the test statistic $\bigvee_{\mathcal{J}_n^\star} w(e)$.

We have a vector $u = (u_p)_{p=1}^N > 0$ of upper bounds on the errors $(e_p)_{p=1}^N$ and a vector of monotonically increasing functions $w = (w_p)_{p=1}^N$. If the total potential margin overstatement $E = \sum_{\mathcal{N}} e$ is big, if $e \leq u$, and if the sample is big enough, it is unlikely that $\bigvee_{\mathcal{J}_n^\star} w(e)$ will be small. So, if the observed value of $\bigvee_{\mathcal{J}_n^\star} w(e)$ is "small enough," that is evidence that $E = \sum_{\mathcal{N}} e < M$— evidence that a full recount would find the same outcome. This section makes the idea precise.

---

[15]If a DRE is working correctly, it should record every vote perfectly. In contrast, if a voter does not use an appropriate pen or pencil to fill in an optically scanned ballot, makes a stray mark on the ballot or does not fill in the bubble perfectly, or if the scanner is miscalibrated, the optical scan could reasonably differ from a human's inference about the voter's intent [Jefferson et al. (2007)].

Let $t \in \mathbb{R}$. Let the sample size $n < N$ be fixed. Define

$$\tag{14} \mathcal{X} = \mathcal{X}(u, M) \equiv \left\{ x \in \mathbb{R}^N : x \leq u \text{ and } \sum_{\mathcal{N}} x \geq M \right\}.$$

The set $\mathcal{X}$ contains all ways of distributing potential margin overstatements across precincts that satisfy the a priori bound $e \leq u$ and the null hypothesis $E \geq M$. To reject the hypothesis $E = \sum_{\mathcal{N}} e \geq M$ when we observe that $\bigvee_{\mathcal{J}_n^\star} w(e) = t$, we need to know that $\mathbb{P}\{\bigvee_{\mathcal{J}_n^\star} w(x) \leq t\}$ is small for all $x \in \mathcal{X}$. Hence, we seek

$$\tag{15} \pi_\star(t) = \pi_\star(t; n, u, w, M) \equiv \max_{x \in \mathcal{X}(u, M)} \mathbb{P}_x \left\{ \bigvee_{\mathcal{J}_n^\star} w(x) \leq t \right\}.$$

The related quantity

$$\tag{16} \pi_\diamond(t) = \pi_\diamond(t; n, u, w, M) \equiv \max_{x \in \mathcal{X}(u, M)} \mathbb{P}_x \left\{ \bigvee_{\mathcal{J}_n^\diamond} w(x) \leq t \right\},$$

where $\mathcal{J}_n^\diamond$ is a random sample of size $n$ *with* replacement from $\mathcal{N}$, is useful to bound the $P$-value when the data come from a stratified sample.

The individual components of $e$ are nuisance parameters: the null hypothesis involves only their sum, $E = \sum_{\mathcal{N}} e$, but the precinct-level potential margin overstatements $\{e_p\}$ affect the probability distribution of $\bigvee_{\mathcal{J}_n^\star} w(e)$, the test statistic.

CLAIM 1. *Let $w^{-1}(t) \equiv (w_p^{-1}(t))_{p=1}^N$. Let $\mathcal{J}_k^-$ be the set of indices of the $k$ smallest components of $u - w^{-1}(t)$. Let $q = q(t, u, w, M)$ be the largest integer for which*

$$\tag{17} \sum_{\mathcal{J}_q^-} u \wedge w^{-1}(t) + \sum_{\mathcal{N} \setminus \mathcal{J}_q^-} u \geq M$$

*or $q = 0$ if there is no such integer. Then*

$$\tag{18} \pi_\star(t; n, u, w, M) = \begin{cases} 0, & q < n, \\ \dfrac{\binom{q}{n}}{\binom{N}{n}}, & q \geq n, \end{cases}$$

*and*

$$\tag{19} \pi_\diamond(t; n, u, w, M) = (q/N)^n.$$

Claim 1 is proved in Appendix A.1.

The following algorithm finds $q$ iteratively:

1. Set $\mathcal{J} = \mathcal{N}$.
2. If $\mathcal{J} = \varnothing$ or $\sum_{\mathcal{J}} u \wedge w^{-1}(t) + \sum_{\mathcal{N} \setminus \mathcal{J}} u \geq M$, $q = \#\mathcal{J}$.
3. Otherwise, let $p \in \mathcal{J}$ attain $[u_p - (u_p \wedge w_p^{-1}(t))] = \bigvee_{\mathcal{J}} [u - (u \wedge w^{-1}(t))]$. (Ties can be broken arbitrarily.) Remove $p$ from $\mathcal{J}$ and return to step 2.

## 4. Putting it together.

4.1. *Testing using a simple random sample of precincts.* Suppose that the precincts for audit will be drawn as a simple random sample. To use the present method, do the following:

1. Select an overall significance level $\alpha$ and a sequence $(\alpha_s)$ so that sequential tests at significance levels $\alpha_1, \alpha_2, \ldots$, give an overall significance level no larger than $\alpha$. For example, we might take $\alpha_s \equiv \alpha/2^s$, $s = 1, 2, \ldots$.
2. Group apparent losing candidates using the pooling rule in Section 3.1.
3. Set the error bounds $u = e^+$.
4. Select a vector of monotonically increasing functions $w = (w_p(\cdot))_{p=1}^N$. For example, $w_p(z) = z$, $w_p(z) = z/b_p$ or $w_p(z) = (z-2)_+/b_p$.
5. Compute the apparent margin $M$.
6. Select an initial sample size $n_1$ and a rule for selecting $n_s$ when the hypothesis $E \geq M$ is not rejected at stage $s - 1$.[16] The only requirement is that $n_1 \geq 0$ and $n_s - n_{s-1} \geq 1$.
7. Set $s = 1$, $n_0 = 0$ and $\mathcal{J}_0 = \varnothing$.
8. Draw a random sample $\mathcal{J}^\star_{n_s - n_{s-1}}$ of size $n_s - n_{s-1}$ from $\mathcal{N} \setminus \mathcal{J}_{s-1}$. Set $\mathcal{J}_s = \mathcal{J}_{s-1} \cup \mathcal{J}^\star_{n_s - n_{s-1}}$. Calculate $\bigvee_{\mathcal{J}_s} w(e)$.
9. If $\pi_\star(\bigvee_{\mathcal{J}_s} w(e); n_s, u, w, M) \leq \alpha_s$, confirm the outcome and stop. Otherwise, increment $s$.
10. If $n_s < N$, return to step 7. Otherwise, audit any precincts not yet in the sample. Confirm the outcome if the outcome was correct.

If there is not a clear set of $f$ winners (if $M = 0$), this will always escalate to a full manual tally.

4.2. *Testing using stratified random samples of precincts.* Under current California law, each county draws its own random sample of 1% of precincts, at a minimum, for post-election audits. (Each county audits at least one precinct for each contest, and fractions are rounded up. Some counties voluntarily audit even larger samples.) Similarly, under Minnesota law, each county draws its own random sample of 2, 3 or 4 precincts for audit, depending on the number of registered voters in the county. The samples in different counties are drawn independently. Thus, for contests that cross county lines, the sample of precincts is a stratified random sample, not a simple random sample. This section presents two ways to combine independent audits of different counties conservatively. Both have merits and shortcomings.

Suppose there are $C$ counties with precincts in the contest. Let $\mathcal{C} \equiv \{1, \ldots, C\}$. Let $E_c$ be the total potential margin overstatement in county $c \in \mathcal{C}$, so $E = \sum_{c \in \mathcal{C}} E_c$. Let $N_c$ be the number of precincts in the contest in county $c \in \mathcal{C}$, so $N = \sum_{c \in \mathcal{C}} N_c$. Let $B_c$ be the number of voting opportunities in the contest in county $c \in \mathcal{C}$, so $B = \sum_{c \in \mathcal{C}} B_c$.

---

[16]Section 4.3 discusses selecting $n_1$. See footnote 12 for approaches to selecting $n_s$.

4.2.1. *Bounds from proportional sampling with replacement.* Fix $n_s > 0$. Let

$$n_{cs} \equiv \lceil n_s N_c / N \rceil. \tag{20}$$

(Note $\sum_{c \in \mathcal{C}} n_{cs} \equiv n'_s \geq n_s$.)

CLAIM 2. Suppose $n_{cs}$ precincts are drawn at random without replacement from county $c$, independently for all $c \in \mathcal{C}$. If there are $k$ precincts among the $N$ in the contest for which $w_p(e_p) \leq t$, the chance that none of the precincts in any of the $C$ samples has $w_p(e_p) > t$ is at most $(k/N)^{n_s}$. This is proved in Appendix A.2.

Essentially, for finding at least one precinct with $w_p(e_p) > t$, stratified sampling without replacement is more effective than stratified sampling with replacement, which is at least as effective as unstratified sampling with replacement if the stratum sample sizes are $\{n_{cs}\}$. So if we draw a sample of size $n_{cs}$ [equation (20)] from county $c$, independently for each $c \in \mathcal{C}$, $\pi_\diamond$ is an upper bound on the maximum $P$-value. This approach computes probabilities as if the sample were drawn with replacement from the entire population of $N$ precincts in the contest, but allocates the sample in proportion to the number of precincts in each county. (Fractions are rounded up, so the actual sample size could be up to $C$ precincts larger than the sample size $n_s$ used in the probability calculations.)

If $N$ is large relative to the overall margin, this method leads to a sample size that is not much larger than required if the sample were a simple random sample from all $N$ precincts in the contest. Each county does a "fair share" of the auditing—the number of precincts a county audits is proportional to the number of precincts in the contest in that county, but for roundoff.

However, if the null hypothesis is not rejected at stage $s$, the sample will need to be expanded in every county in the contest (but for roundoff). Moreover, whether such an expansion is needed depends on the audit results from all counties in the contest, so county audit schedules are interdependent. In contrast, the approach in the next subsection typically requires auditing more precincts, but the audits in different counties are logistically independent: whether the audit in a given county needs to be expanded depends on the audit results in that county alone.

Stanislevic (2006) makes a claim that implies that the probability that none of the precincts in the stratified sample has $w_p(e_p) > t$ is at most $\binom{k}{n_s}/\binom{N}{n_s}$, that is, stratification using sample sizes $\{n_{cs}\}$ can only help. If that conjecture were true, one could calculate the maximum $P$-value using $\pi_\star$ instead of $\pi_\diamond$, and the sample size would be at most $C$ precincts larger than that required for a simple random sample from the $N$ precincts in the contest. The conjecture is false, but seems to be "almost true."[17]

Note that this approach can be used to find a conservative $P$-value for any set of sample sizes $n_{cs}$ by pretending that the overall sample size corresponds to the smallest sampling fraction $n_{cs}/N_c$; that is, that the data came from a sample of size $n_s = \lfloor N \bigwedge_{c \in \mathcal{C}} (n_{cs}/N_c) \rfloor$ drawn with replacement from the population of $N$ precincts. If the sampling fractions vary widely by county, this can be extremely conservative. See Section 5.2 for an illustration.

4.2.2. *Bounds from independent tests in every county.*

CLAIM 3. There must be at least one county $c \in \mathcal{C}$ for which $\frac{E_c}{B_c} \geq \frac{E}{B}$. This is proved in Appendix A.3.

Suppose we test in each county at significance level $\alpha$ whether $E_c \geq MB_c/B$. Let $R_c$ be the event that the test in county $c$ rejects the hypothesis $E_c \geq MB_c/B$. Then, if in at least one county $E_c \geq MB_c/B$,

$$\Pr\left(\bigcap_{c \in \mathcal{C}} R_c\right) \leq \bigwedge_{c \in \mathcal{C}} \Pr(R_c) \leq \alpha \tag{22}$$

---

[17]Stanislevic [personal communication (2007)] notes that there are counterexamples, but his numerical experiments suggest that increasing $n$ by one restores the inequality. Moreover, he claims that the inequality fails only when the counties have equal size and $k$ and $n_s$ are divisible by $C$, and that when the inequality fails,

$$\frac{\binom{k}{n}}{\binom{N}{n}} < \left(\frac{\binom{k/C}{n/C}}{\binom{N/C}{n/C}}\right)^C. \tag{21}$$

That is, taint is hardest to detect when the counties are the same size and have the same number of tainted precincts. Here is an example: Take $N = 100$, $n_s = 80$, $k = 98$, $C = 2$, $N_1 = N_2 = 50$, $n_{1s} = n_{2s} = \lceil 80/2 \rceil = 40$, $k_1 = k_2 = 49$ (i.e., one heavily tainted precinct in each county). Then the chance a simple random sample of size 80 from the 100 precincts contains neither of the two heavily tainted precincts is $\frac{\binom{98}{80}}{\binom{100}{80}} = 3.8\%$, but the chance that a stratified random sample that draws 40 precincts from each of the two counties without replacement contains neither of the two heavily tainted precincts is $\left(\frac{\binom{49}{40}}{\binom{50}{40}}\right)^2 = 4\%$. In this case, the chance of finding a heavily tainted precinct is less for the stratified random sample than for the simple random sample: stratification can hurt. (If both heavily tainted precincts are in the same county, stratification helps.) If $n_s$ is increased to 81 so that $n_{cs} = 41$ precincts are drawn from each county, then stratification helps. The situation with stratification is rather delicate.

(the probability of an intersection of events is no greater than the smallest of the event probabilities). Thus, if the total error $E$ across precincts is $M$ or greater, the chance that we conclude at significance level $\alpha$ that $E_c < MB_c/B$ in every one of the $C$ counties is at most $\alpha$ overall, and typically rather less.[18]

This approach can be quite conservative. When the counties all contain many precincts in the contest and the margin is large, the overall sample size will tend to be about $C$ times larger than would be required if the sample were drawn without stratification. This wastes resources.

However, the approach has some logistical advantages. The apparent margin depends on results in every county involved in the contest, so there must be communication among counties before the audit can begin. But, unlike the previous method, errors detected in one county do not require any other county to increase its sample size, and the audit process does not require cooperation or communication among counties.

4.3. *"Fault-tolerant" initial sample size.* The procedure can start with any initial sample size $n_1 \geq 0$. However, if the initial sample size $n_1$ is too small, we will not be able to reject the hypothesis $E \geq M$ on the basis of the initial sample even if it shows no miscount whatsoever. Audit samples often show small miscounts.

We can determine an initial sample size $n_1$ so that we can confirm the outcome without expanding the sample, provided the potential margin overstatement found in the initial sample is sufficiently small. For example, suppose we would like to be able to confirm the outcome as long as the test statistic evaluated for the initial sample is no greater than $t_1$. If we choose

$$(23) \qquad n_1 = \arg\min_{n>0}\{n : \pi_\star(t_1, n, u, w, M) < \alpha_1\},$$

then if $\bigvee_{\mathcal{J}^\star_{n_1}} w(e) \leq t_1$, we can confirm the outcome without expanding the sample. Section 5.1 gives an example of this calculation.

If we are drawing a stratified random sample, we need to find an initial sample size $n_{1c}$ for each county $c$. For the approach to stratification in Section 4.2.1, we can take $n_{1c} = \lceil n_1 N_c/N \rceil$, with

$$(24) \qquad n_1 = \arg\min_{n>0}\{n : \pi_\diamond(t_1, n, u, w, M) < \alpha_1\}.$$

---

[18]Dopp and Stenger have asserted that to audit contests that span more than one county, one should set the sample size using the smaller of the county or state margin [Dopp and Stenger (2006)]. To the best of my knowledge, they have not investigated the effect that has on confidence in the outcome of the election, and gave no proof that it results in a conservative test. This proof shows that if one uses the overall margin—scaled by the number of ballots voted in the contest in the county in question—the result is conservative.

For the approach to stratification in Section 4.2.2, the calculation is more complex. Let $u_c$ denote the vector of precinct error bounds for county $c$ and let $w_c$ denote the vector of precinct weight functions for county $c$. If the initial sample size for county $c$ is chosen to be

$$(25) \qquad n_{1c} = \arg\min_{n>0}\{n : \pi_\star(t_1, n, u_c, w_c, \lfloor MB_c/B \rfloor) < \alpha_1\},$$

we will not have to expand the audit in county $c$, provided the test statistic for the initial sample is no greater than $t_1$.

**5. Examples.** This section gives examples of calculating $P$-values for the hypothesis that the apparent outcome of an election is wrong. It does not give examples of expanding the sample size sequentially: data required for those computations are not available.

5.1. *November 2006 Sausalito Marin City school board race.* The November 2006 school board race for the Sausalito Marin City School District in Marin County, California involved nine precincts. Voters could vote for three of five candidates or a write-in. Table 2 lists vote totals by precinct for each candidate. Absentee and polling-place votes were combined.

The winning candidate with the fewest votes was Mark Trotter, with 2022 votes. The losing candidate with the most votes was George Stratigos, with 1936 votes. The margin between the two was $2022 - 1936 = 86$ votes—an extremely narrow margin of 0.57% of the 15,000 possible votes. If 43 votes for Stratigos had been awarded erroneously to Trotter, that would have sufficed to change a tie (1979 votes each) into a win for Trotter, with a net potential margin overstatement in the election of 86. Any other change to the set of winners would have required a larger potential margin overstatement. Thus, if we can reject the hypothesis that the total potential margin overstatement is greater than or equal to 86 votes, we can conclude that the outcome of the election was correct.

Every unexercised opportunity to vote counts as an undervote. In this example, a ballot can contribute up to three undervotes: the number of undervotes on a ballot is 3 minus the number of candidates voted for, provided the number voted for is no greater than three. If a voter marked the ballot for more than three candidates, the ballot contributes overvotes.

We shall take $w_p(z) = z/b_p$, so that the test statistic is the maximum potential margin overstatement as a fraction of the voting opportunities in each precinct in the sample. Note that the number of votes for write-ins plus the number of votes for Peter C. Romanowsky is less than the number of votes for the runner-up, George T. Stratigos, but the number of undervotes plus three times the number of invalid ballots is greater than the number of votes for Stratigos. Therefore, write-ins can be pooled with each other and with Romanowsky, but undervotes and invalid ballots are treated as a separate candidate, as described in Section 3.1.

Table 2
*Vote totals by precinct for the November 2006 Sausalito Marin City School Board race. Voters could vote for up to three candidates. The number of undervotes is three times the number of ballots, minus the total number of votes for candidates, ignoring ballots showing votes for more than three candidates (overvoted ballots). Column 9, "votes," is the total number of voting opportunities, three times the number of ballots. There were 5,000 ballots, including two overvoted ballots, one in precinct 3104 and one in precinct 3601. The post-election audit examined all the ballots in precinct 3107 and found a discrepancy of one vote. The discrepancy was due to operator error; re-scanning the ballots eliminated the discrepancy. [E. Ginnold, Registrar of Voters, Marin County, California, personal communication (2007).] Data courtesy of E. Ginnold and M. Briones*

| Precinct | Undervotes + 3 × overvotes | Thornton | Hoyt | Trotter | Stratigos | Romanowsky | Write-ins | Votes |
|---|---|---|---|---|---|---|---|---|
| 3001 | 780 | 296 | 309 | 283 | 271 | 60 | 5 | 2004 |
| 3002 | 920 | 311 | 287 | 274 | 291 | 44 | 3 | 2130 |
| 3104 | 699 | 238 | 244 | 240 | 225 | 48 | 4 | 1698 |
| 3105 | 765 | 270 | 262 | 240 | 228 | 56 | 3 | 1824 |
| 3106 | 668 | 239 | 267 | 294 | 209 | 58 | 5 | 1740 |
| 3107 | 732 | 251 | 260 | 236 | 214 | 53 | 3 | 1749 |
| 3600 | 582 | 235 | 233 | 129 | 186 | 51 | 6 | 1422 |
| 3601 | 367 | 234 | 178 | 126 | 170 | 40 | 7 | 1122 |
| 3602 | 610 | 160 | 155 | 200 | 142 | 39 | 5 | 1311 |
| Total | 6123 | 2234 | 2195 | 2022 | 1936 | 449 | 41 | 15,000 |

Table 3 gives three potential margin overstatement bounds: the a priori bounds $e^+$ based on pooling write-in candidates only, $e^+$ based on pooling write-in candidates and Romanowsky, and $\lceil 0.4b \rceil$, 40% of the votes, including undervotes and three times the overvotes, rounded up to the next integer. For all three bounds, any of the nine precincts could harbor enough miscount to change the apparent outcome of the election.

Suppose we want to design an initial sample size so that, provided the maximum potential margin overstatement in any precinct in the sample is no more than 0.2% of the votes reported in that precinct (including undervotes and three times the overvotes), we would reject the hypothesis that the wrong set of winners was named at significance level 0.01 (we would confirm the outcome at "confidence level" 99%). That corresponds to rejecting the hypothesis when $\bigvee_{\mathcal{J}_n^\star} w(e) \leq 0.002$. Note that $0.002 \times 15{,}000 = 30 < 86$, so at least one precinct must have more than this background level of error (0.2%) for the outcome of the election to be wrong.

TABLE 3
*Three possible bounds on the potential margin overstatement in each precinct. The bound $e^+(b)$ is defined in equation ([7](#)). Column 2 pools the write-in candidates in computing $e^+$. Column 3 pools the write-in candidates and Peter C. Romanowsky in computing $e^+$, which leads to smaller bounds on the error; see Section [3.1](#). The bound $\lceil 0.4b \rceil$ is 40% of the reported voting opportunities in the precinct, rounded up to the next integer. This is analogous to the maximum within-precinct error bounds used by Saltman (1975), Dopp and Stenger (2006) and McCarthy et al. (2008)*

| Precinct | $e^+(b)$ write-ins pooled | $e^+(b)$ write-ins & Romanowsky pooled | $\lceil 0.4b \rceil$ |
|---|---|---|---|
| 3001 | 2887 | 2827 | 802 |
| 3002 | 2999 | 2955 | 852 |
| 3104 | 2416 | 2368 | 680 |
| 3105 | 2593 | 2537 | 730 |
| 3106 | 2535 | 2477 | 696 |
| 3107 | 2493 | 2440 | 700 |
| 3600 | 2013 | 1962 | 569 |
| 3601 | 1653 | 1613 | 449 |
| 3602 | 1821 | 1782 | 525 |

Thus,

$$\pi_\star(0.002, n, u, w, 86) = \frac{\binom{8}{n}}{\binom{9}{n}}. \tag{26}$$

Enough miscount to change the outcome could lurk in a single precinct. Suppose that just one precinct had miscount, and that the miscount was enough to change the outcome of the election. Then even if we audited 8 of the 9 precincts at random, there is a one-in-nine chance that we would fail to audit that precinct. So, to have 99% confidence in the outcome of this race if the observed potential margin overstatement were at most 0.2% of the votes (including undervotes and overvotes) in any precinct, we would have to audit every precinct. That is bad news, but since the margin is only 0.57% of the possible votes, it is not surprising.

In fact, one precinct was audited (precinct 3107) and it was found to contain one error. We shall presume that this error favored one of the apparent winners. The number of votes in precinct 3107 is 1749, so this corresponds to a test statistic value $\bigvee_{\mathcal{J}_1^\star} w(e) = 1/1749 = 0.00057$. On the basis of this audit, the maximum $P$-value of the hypothesis that the wrong set of three candidates was declared the winner is

$$\pi_\star(0.00057, 1, u, w, 86) = \frac{\binom{8}{1}}{\binom{9}{1}} = 88.9\%. \tag{27}$$

So, even if there were enough miscount in the aggregate to cause the apparent set of winners to differ from the true set of winners, the chance that an audit of one precinct would show $w_p(e_p) \leq 0.00057$ could be as large as 88.9%, depending on how the miscount is distributed across precincts.

Conversely, what would we have to believe about the error for these audit data to yield a $P$-value of 1% or less? For a random sample of just one of the nine precincts to have at most a 1% chance of having $\bigvee_{\mathcal{J}_1^\star} w(e) \leq 0.00057$, all nine precincts would have to have $w_p(e_p) > 0.00057$. For an election-altering discrepancy to require $w_p(e_p) > 0.00057$ in every precinct corresponds to a bound $u = 0.0057b$. Unless we believe that a potential margin overstatement of more than 0.0057% of the votes is either impossible or certain to be detected without an audit, we could not possibly get 99% confidence in the outcome of this race by auditing only one precinct.

5.2. *November 2006 Minnesota U.S. Senate race.* This section examines the November 2006 Senate race in Minnesota. Minnesota has 87 counties with a total of 4,123 precincts, of which 202 were audited after the election. Table 4 lists the vote totals for the race. The winner was Amy Klobuchar and the runner-up was Mark Kennedy. The statewide margin of victory was 443,196 votes for 2,217,818 voters, 20.0% of voters (not of cast votes).[19]

The audit of this election is discussed by Halvorson and Wolff (2007). Minnesota elections law S.F. 2743 (2006) requires auditing a random sample of precincts in each county, with a sample size that depends on the voting population in the county: counties with fewer than 50,000 registered voters must audit at least two precincts; counties with between 50,000 and 100,000 registered voters must audit at least three; and counties with more than 100,000 registered voters must audit at least four precincts. At least one of the precincts audited in each county must have 150 or more votes cast. Hennepin County audited eight precincts instead of the four required. (It still had the smallest sampling fraction.) Several other counties also audited more than the minimum required.

TABLE 4
*Summary of 2006 U.S. Senate race in Minnesota*

| Voters | Undervotes & invalid ballots | Fitzgerald (Indep) | Kennedy (Repub) | Klobuchar (Democ/Farm/ Labor) | Cavlan (Green) | Powers (Constit) | Write-ins |
|---|---|---|---|---|---|---|---|
| 2,217,818 | 15,099 | 71,194 | 835,653 | 1,278,849 | 10,714 | 5,408 | 901 |

---

[19]Data in this section come from www.sos.state.mn.us/docs/2006_General_Results.XLS, electionresults.sos.state.mn.us/20061107/ElecRslts.asp?M=S&Races=0102 and www.sos.state.mn.us/home/index.asp?page=544.

Precincts audited had from 2 to 2,393 ballots cast.[20] The largest value of $e_p$ was 2; the largest value of $e_p/b_p$ was 0.67%. The total observed discrepancy was 62 votes, about 0.065% of ballots cast in the audited precincts, including undervotes and invalid votes. The total observed potential margin overstatement was 25 votes, about 0.026% of ballots.

The audit shows a different number of ballots from that reported in 13 of the audited precincts: "ballot accounting" apparently had not been done. Ten of the differences were one ballot each. In two precincts,[21] the number of ballots was off by three. Most of the discrepancies in vote totals seem to have been caused by jams in the optical scanner or by ballots fed through the scanner twice. The observed discrepancies in $b_p$ are not large enough to affect the error bounds $e^+$ or $0.4b$ by much, but they show that $b_p$ is not an inviolable upper bound on $a_p$, and there might be larger discrepancies in the precincts not sampled.

Under Minnesota law, auditors can interpret voter intent, even if the ballot is not marked properly.[22] In one precinct,[23] three machine-unreadable ballots originally tallied as undervotes were interpreted by the auditors as votes for Amy Klobuchar. The precinct had only 96 voters, so a three-vote error is a large percentage of $b_p$—although in this case the error does not contribute to $e_p$ because it favors Klobuchar, the winner. This illustrates why taking $w_p(z) = z/b_p$ is perhaps too sensitive to occasional errors, and $w_p(z) = z$ or $w_p(z) = (z-m)_+/b_p$ might be preferable.

We will calculate $P$-values for the hypothesis that a full manual recount would not find that Amy Klobuchar is the winner, under a variety of assumptions. Because the Minnesota law links sample sizes to the number of registered voters in each county rather than to the number of precincts in each county and never requires more than 4 precincts per county, the sampling fraction of precincts varies widely from county to county. The minimum sampling fraction in the 2006 audit was 1.9% and the maximum was 23.8%. Two-thirds of the counties had precinct sampling fractions between 4% and 9%. Only one had a sampling fraction below 2%—the largest county, Hennepin. The overall sampling fraction was 4.9% of precincts.

Reported undervotes, overvotes and votes for all other candidates total less than the vote reported for runner-up Mark Kennedy, so they can all be pooled into one pseudo-candidate as described in section 3.1. Thus, we have $K = 3$ pseudo-candidates, $f = 1$, $N = 4{,}123$, $B = 2{,}217{,}818$, $M = 443{,}196$. We will consider two upper bounds on the precinct-level miscount, $u = e^+(b)$ and $u = 0.4b$, and three functions for weighting the precinct-level potential margin overstatements, $w_p(z) = z$, $w_p(z) = z/b_p$ and $w_p(z) = (z-2)_+/b_p$.

---

[20]Mean 471, median 272, IQR 505.

[21]Spring Lake Park Precinct 3 and Orono Precinct 2.

[22]However, discrepancies caused by machine-unreadable ballots do not trigger an escalation of the audit.

[23]Lee Township, Norman County.

TABLE 5
*The smallest number of precincts in Minnesota as a whole that must have $w_p(e_p) > \bigvee_{\mathcal{J}} w(e)$ for the outcome of the election to differ from the outcome a full manual recount would show, where $\mathcal{J}$ is the set of indices of precincts actually sampled. Here $\bigvee_{\mathcal{J}} w(e)$ is the observed value of the test statistic for the 202 precincts in the sample. Values are given for three choices of the weight functions $w_p$ and two bounds $u$ on the amount of error each precinct can hold*

|              | $w_p(z) = z$ | $w_p(z) = z/b_p$ | $w_p(z) = (z-2)_+/b_p$ |
|---|---|---|---|
| $u = e^+(b)$ | 130 | 128 | 130 |
| $u = 0.4b$   | 721 | 720 | 721 |

The approach to dealing with stratification in Section 4.2.2 leads to very large $P$-values in this example—over 27% for all six combinations of $u$ and $w_p$. We can get a very conservative $P$-value by pretending that the sample was drawn with replacement from the entire population of precincts, but that only 1.9% of the precincts (78) were sampled; this is an application of the bound in Section 4.2.1. Table 5 shows the lower bounds on the number of precincts statewide that would have to have potential margin overstatements greater than $w_p^{-1}(\bigvee_{\mathcal{J}} w(e))$ in order to have $E \geq M$ (here $\mathcal{J}$ are the indices of the 202 precincts in the actual sample).

Table 6 gives the corresponding $P$-values. It also gives $P$-values using the same observed discrepancies, but pretending that the sample of 202 precincts was drawn in two other ways: as a stratified sample with sample size proportional to the number of precincts in each county, using the bound derived in Section 4.2.1, or as a simple random sample of 202 precincts. Had the 202 precincts been drawn in either of those ways, the $P$-values would be much smaller than the bound derived for the sampling scheme Minnesota actually used.

Table 6 shows that the audit data would allow us to reject the hypothesis that a full recount would find a different winner at significance level 10%, for all three choices of test statistics and for either error bound. Stark (2008b) finds $P$-values about half as large using a sharper measure of discrepancy. For the error bound $u = 0.4b$, we could reject the hypothesis at significance level 1%. If the data had come from a simple random sample from the state as a whole, or if the sample size in each county had been proportional to the number of precincts in the county, we would have been able to reject the hypothesis that the apparent outcome differs from the outcome a full manual recount would show at significance level 1%.

CONSERVATIVE ELECTION AUDITS 27

Table 6

*P-values for the hypothesis that a full manual recount would show that Amy Klobuchar did not win the Senate race, under different assumptions about how the sample was drawn and the potential margin overstatement in each precinct [upper bounds $u = e^+(b)$ and $u = 0.4b$], and different choices of the weighting of errors in each precinct. The first row is for precinct-level weight function $w_p(z) = z$: each error has the same weight. The second row is for $w_p(z) = z/b_p$: errors in larger precincts have lower weight. The third row is for $w_p(z) = (z-2)_+/b_p$: that test statistic ignores the first two potential margin overstatements in each precinct; after the first two, potential margin overstatements in larger precincts have lower weight. Columns 2 and 3 are very conservative upper bounds derived by treating the sample as if it were a smaller sample of 1.9% of the precincts in each county (78 precincts in all, rather than 202). Columns 4 and 5 pretend that the data came from a stratified random sample of 202 precincts in which the number of precincts drawn from each county is proportional to the number of precincts in the county. Columns 6 and 7 pretend that the data came from a simple random sample from all the precincts in the state. Only the results in columns 2 and 3 apply to the auditing scheme Minnesota actually used*

| | 1.9% sample w/ replacement | | Proportional sample | | Sample w/o replacement | |
|---|---|---|---|---|---|---|
| | $u = e^+(b)$ | $u = 0.4b$ | $u = e^+(b)$ | $u = 0.4b$ | $u = e^+(b)$ | $u = 0.4b$ |
| $w_p(z) = z$ | 8.2% | 0.00003% | 0.15% | $1.4 \times 10^{-15}$% | 0.13% | $4.6 \times 10^{-16}$% |
| $w_p(z) = z/b_p$ | 8.5% | 0.00003% | 0.17% | $1.5 \times 10^{-13}$% | 0.15% | $4.9 \times 10^{-16}$% |
| $w_p(z) = (z-2)_+/b_p$ | 8.2% | 0.00003% | 0.15% | $1.4 \times 10^{-15}$% | 0.13% | $4.6 \times 10^{-16}$% |

## 6. Discussion.

6.1. *P-values.* As illustrated in Section 5, the method can also find the maximum P-value of the hypothesis that $E \geq M$ and hence of the hypothesis that the election outcome is incorrect given discrepancy data from a particular sampling design. The maximum P-value is $\pi_\star(\bigvee_{\mathcal{J}_1} w(e); n_1, u, w, M)$, where $\mathcal{J}_1$ is the initial random sample, of size $n_1$. This expression applies only to the initial sample. If the approach is used sequentially, the P-values need to be adjusted to take that into account.

6.2. *Two-position contests requiring super-majority.* The bound $e^+$ on potential margin overstatement can be sharpened easily for contests such as ballot measures or propositions that have only two positions and that require more than a simple majority to pass. For example, suppose that a contest allows only "yes" or "no" votes, and requires a 2/3 majority of "yes" votes to pass. Suppose that, according to the reported totals, the measure passed. Let "yes" be candidate $k = 1$ and "no" be candidate $k = 2$. The effective apparent margin is the margin above 2/3 of the total vote:

$$(28) \qquad M = \lfloor V_1 - \tfrac{2}{3}(V_1 + V_2) \rfloor.$$

An error that increases $V_1$ by one vote increases $V_1 - \frac{2}{3}(V_1 + V_2)$ by only $1/3$ of a vote. An error that decreases $V_2$ by one vote increases $V_1 - \frac{2}{3}(V_1 + V_2)$ by $2/3$ of a vote. Within each precinct, error could have inflated the effective apparent margin over $2/3$ by no more than

$$(29) \qquad \lceil (v_{1p} - \tfrac{2}{3}(v_{1p} + v_{2p})) + \tfrac{2}{3}r_p \rceil = \lceil \tfrac{2}{3}(r_p + v_{1p}/2 - v_{2p}) \rceil.$$

These are smaller upper bounds $u$ for $e$ than $e^+$ are, but still rigorous.

6.3. *Why not use the sample sum or sample mean?* Using the discrepancy of the totals across the precincts in the sample as the test statistic instead of calculating the discrepancy separately for each precinct would have advantages. For example, it would allow errors that hurt a particular candidate to cancel errors favoring that candidate in a different precinct, which might allow us to reject the hypothesis that the wrong candidate was named the winner using smaller samples. However, it is far more difficult to calculate tail probabilities for the discrepancy of the totals. In particular, it is not true that the most difficult-to-detect election-altering taint concentrates as much miscount as possible in as few precincts as possible, precisely because cancellations can occur.

6.4. *Improving the power.* The approach presented here is conservative: the chance that it declares the outcome to be correct when the outcome is not correct is at most $\alpha$. However, other approaches could do the same thing using smaller audit samples—they could have more power for the same significance level.

The elements of the approach with the most room for improvement are these:

1. The test statistic, pooling and aggregation of the miscount. There are sharper necessary conditions and measures of discrepancy; see, for example, Stark (2008b). The functions $\{w_p\}_{p \in \mathcal{N}}$ could be optimized against various alternatives. One could construct a more powerful test using likelihood ratios or the sample sum, as described in Section 6.3. However, these improvements in power come at a cost of far more complex probability calculations and a loss of transparency to jurisdictional users. Numerical optimization would appear to be necessary to calculate $P$-values.
2. Stratification. The approaches to dealing with stratification for contests that cross county lines are conservative but not sharp. Better inequalities would allow smaller samples to be used.
3. Thresholds for sequential tests. The inequalities used to set the significance levels in the sequential tests could be improved.

4. Sample design. If we were at liberty to choose the sampling design, a different approach—such as sampling with probability proportional to $u_p$, the upper bound on the potential margin overstatement—might permit smaller samples.[24]

Ideas from sequential analysis [Siegmund (1985) and Wald (2004)] could certainly help improve the thresholds for sequential testing.

6.5. *Alternative approaches.* One could also take a Bayesian approach to the problem: given a prior probability distribution, one could compute posterior odds that the election named the right winner given the audit data, and confirm the outcome if those odds were, say, 100 to 1 or greater. This approach requires prior probability distributions for the number of votes for each candidate and for the potential margin overstatement.

The false discovery rate [Benjamini and Hochberg (1995)] gives another perspective: rather than insist that the chance of confirming an outcome that is incorrect be no larger than $\alpha$, we could require the expected fraction of confirmed election outcomes that are confirmed in error to be no larger than $\alpha$.

A rather different approach is to combine a base rate of random sampling with "targeted" sampling, where candidates or other interested parties select some precincts for audit by any means they choose [Norden et al. (2007) and Jefferson et al. (2007)]. Computing a $P$-value for this approach would require an ad hoc model for the efficacy of "educated guesses" in finding miscounted precincts, but the method could increase public confidence in the election outcome.

Any of these approaches is incomplete without rules for expanding the audit if the precincts in the targeted sample show material miscount, culminating either in confirming the outcome or in a full recount.

**7. Conclusions.** Post-election audits can be used to confirm election outcomes or show that a full manual recount is needed. The election outcome is confirmed if, on the assumption that the election outcome is incorrect, the probability is large that the sample would have contained larger potential margin overstatements than it did contain. If that probability is not sufficiently large, the sample size needs to be increased. Eventually, either the sample includes every precinct (there has been a complete manual recount), or there is compelling statistical evidence that the election outcome is correct.

---

[24]See, for example, Aslam, Popa and Rivest (2007) and Stark (2008a). Current and pending audit laws do not contemplate sampling designs other than simple or stratified random samples. If sampling with probability proportional to $u_p$ were allowed, it would bring election auditing much closer to work in financial auditing, where *monetary unit sampling* is often used [Panel on Nonstandard Mixtures of Distributions (1989)]. However, if precincts have differing probabilities of selection, so do ballots, which might raise legal issues of differential enfranchisement.

Confirming an election outcome statistically requires upper bounds on the potential margin overstatement in each precinct. Such upper bounds can be calculated from upper bounds on the total number of votes in each precinct. Upper bounds on the number of votes could in turn come from the number of registered voters, from the number of ballots issued to precincts, from precinct pollbooks or from "ballot accounting." Alternatively, one might use ad hoc bounds on the potential margin overstatement, such as 40% of the number of reported ballots in the precinct. The results are sensitive to the bounds, so ad hoc choices need to be justified and tested empirically in every election.

Combining a base rate of sampling (such as California's 1% law) with rules for increasing the sample size for contests, where—given the margin, the number of ballots cast in each precinct and the miscount observed in the initial sample—the outcome is in doubt, is a statistically sound and potentially practical way[25] to use post-election audits to decide whether to confirm the outcome. The base rate of sampling provides a broad check for gross errors; increasing the sample size for close contests and contests where the audit reveals potential margin overstatements can guarantee any desired level of confidence in the outcome.

In states where election regulations do not contemplate increasing the size of an initial audit, the approach outlined here can be used to calculate the confidence that each election outcome is correct,[26] given the size of the sample, the margin, the reported votes in each precincts and the potential margin overstatements observed in the sample.

## APPENDIX

**A.1. Proof of Claim 1.** Both $\mathbb{P}_x\{\bigvee_{\mathcal{J}_n^\star} w(x) \leq t\}$ and $\mathbb{P}_x\{\bigvee_{\mathcal{J}_n^\diamond} w(x) \leq t\}$ are monotonic in $\#\{p : x_p \leq w_p^{-1}(t)\}$. Hence, $\pi_\star(t)$ and $\pi_\diamond(t)$ are attained by the element $x^-$ of $\mathcal{X}$ with the fewest components greater than the corresponding components of $w^{-1}(t)$. To maximize $\#\{p : x_p \leq w_p^{-1}(t)\}$ while keeping $E = \sum_\mathcal{N} x \geq M$ and $x \leq u$, set $x_p = u_p$ for those components $p$ for which $u_p - w_p^{-1}(t)$ is largest, and set the remaining components of $x$ to whichever is smaller, $u_p$ or $w_p^{-1}(t)$. Thus, for some $k$, $x^-$ is of the form

$$
(30) \quad x_p^- = \begin{cases} (u \wedge w^{-1}(t))_p, & p \in \mathcal{J}_k^- \\ u_p, & p \notin \mathcal{J}_k^-. \end{cases}
$$

---

[25]The method was tested in practice in Marin County, California, to audit Measure A on the 5 February 2008 ballot to attain 75% confidence that a full manual count would match the apparent outcome.

[26]As mentioned above, "confidence" that the outcome is correct is taken to mean 100% minus the $P$-value of the hypothesis that the outcome is incorrect; this is not a standard statistical definition of "confidence."

The value of $k$ that gives $x^-$ is the largest possible value for which $E \geq M$, namely, $q$ [defined in equation (17)]. The chance that $\mathcal{J}_n^\star(w(x^-)) \leq t$ is the chance that $\mathcal{J}_n^\star$ consists of $n$ of the $q$ components of $x^-$ that are less than the corresponding components of $w^{-1}(t)$, as equation (18) asserts. Similarly, the chance that $\mathcal{J}_n^\diamond(w(x^-)) \leq t$ is the chance that $\mathcal{J}_n^\diamond$ includes only components $x^-$ that are less than the corresponding components of $w^{-1}(t)$. There are $q$ such components, so the chance is $(q/N)^n$, as claimed.

**A.2. Proof of Claim 2.** Among the $N$ precincts in the contest, $k$ have $w_p(e_p) \leq t$. We divide the $N$ precincts into $C$ strata. In stratum $c$, there are $N_c$ precincts of which $k_c$ precincts have $w_p(e_p) \leq t$, and $\sum_{c \in \mathcal{C}} k_c = k$. We draw $n_{cs} = \lceil n_s N_c / N \rceil$ precincts at random without replacement from county $c$. Let $n_s' = \sum_{c \in \mathcal{C}} n_{cs} \geq n_s$. Let $S_c$ be the number of precincts in the sample from county $c$ for which $w_p(e_p) > t$. Then $S_c$ has the hypergeometric distribution with parameters $N_c$, $N_c - k_c$ and $n_{cs}$, and $\{S_c\}_{c \in \mathcal{C}}$ are independent. Moreover,

$$
(31) \quad \mathbb{P}\{S_c = 0\} = \frac{\binom{k_c}{n_{cs}}}{\binom{N_c}{n_{cs}}} \leq (k_c/N_c)^{n_{cs}}.
$$

This follows from the fact that $\frac{x}{y} > \frac{x-1}{y-1}$ when $x < y$ and $y > 1$. Because the samples from different strata are independent,

$$
(32) \quad \mathbb{P}\left\{\sum_{c \in \mathcal{C}} S_c = 0\right\} \leq \prod_{c \in \mathcal{C}} (k_c/N_c)^{n_{cs}}.
$$

Since $n_{cs} \geq n_s N_c / N$, and $n_s = \sum_{c \in \mathcal{C}} n_s N_c / N$,

$$
\prod_{c \in \mathcal{C}} (k_c/N_c)^{n_{cs}} \leq \prod_{c \in \mathcal{C}} (k_c/N_c)^{n_s N_c / N}
$$

$$
(33) \quad \leq \left(\frac{1}{n_s} \sum_{c \in \mathcal{C}} (n_s N_c / N)(k_c/N_c)\right)^{n_s}
$$

$$
= (k/N)^{n_s}.
$$

The second step is an application of the arithmetic mean–geometric mean inequality. Hoeffding (1956), Theorem 4, proves something rather more general.

Inequality (33) shows that if we draw a stratified sample of precincts with $n_{cs}$ precincts from county $c$, $c \in \mathcal{C}$, but compute the maximum $P$-value as if we were sampling *with* replacement from the entire population of $N$ precincts (i.e., if we use $\pi_\diamond$ as the bound on the $P$-value), we get a conservative test.

**A.3. Proof of Claim 3.** Claim 3 just asserts that either every element of a list is equal to the mean of the list, or there is at least one element greater than the mean:

$$
\begin{aligned}
\frac{E}{B} &= \frac{\sum_{c \in \mathcal{C}} E_c}{B} \\
&= \frac{\sum_{c \in \mathcal{C}} B_c (E_c/B_c)}{B} \\
&= \sum_{c \in \mathcal{C}} B_c/B (E_c/B_c) \\
&\leq \left( \sum_{c \in \mathcal{C}} B_c/B \right) \times \bigvee_{c \in \mathcal{C}} \frac{|E_c|}{B_c} \\
&= B/B \times \bigvee_{c \in \mathcal{C}} \frac{|E_c|}{B_c} \\
&= \bigvee_{c \in \mathcal{C}} \frac{|E_c|}{B_c}.
\end{aligned}
\tag{34}
$$

The antepenultimate step follows from Hölder's inequality.

So, if the total potential margin overstatement $E$ across counties is $M$ or more, there must be at least one county $c$ for which $E_c \geq MB_c/B$.

**Acknowledgments.** I am grateful to Vittorio Addono, Kim Alexander, Alessandra Baniel-Stark, Kathy Dopp, Stephen Fienberg, David Freedman, Joe Hall, Mark Halvorson, David Jefferson, Mark Lindeman, John McCarthy, Jasjeet Sekhon, Howard Stanislevic, David Wagner and an anonymous referee for helpful conversations and comments on an earlier draft, and to Elaine Ginnold and Melvin Briones for data.

Department of Statistics
University of California
Berkeley, California 94720-3860
USA
E-mail: stark@stat.berkeley.edu